\documentclass{moriond}

\usepackage{amsmath}

\def\Journal#1#2#3#4{{#1} {\bf #2}, #3 (#4)}

\def\be{\begin{equation}}
\def\ee{\end{equation}}
\def\bea{\begin{eqnarray}}
\def\eea{\end{eqnarray}}

\vspace{4cm}
\title{MEASUREMENT OF THE GRAVITATIONAL REDSHIFT EFFECT WITH THE RADIOASTRON SATELLITE}

\author{A.V. BIRIUKOV $^1$, V.L. KAUTS $^{1,2}$ ,D.A. LITVINOV $^3$, N.K. PORAYKO $^3$, V.N. RUDENKO $^3$}
\author{$^1$ Astro Space Center, Lebedev Physical Institute, Moscow, Russia\\
$^2$ Bauman Moscow State Technical University, Moscow, Russia\\
$^3$ Sternberg Astronomical Institute, Moscow State University, Moscow, Russia
\vspace{12pt}}

\begin{document}

\maketitle

\abstracts{RadioAstron satellite admits in principle testing the gravitational redshift effect with an accuracy of better than $10^{-5}$. It would surpass the result of Gravity Probe A mission at least by an order of magnitude. However, RadioAstron's communications and frequency transfer systems are not adapted for a direct application of the non relativistic Doppler and troposphere compensation scheme used in the Gravity Probe A experiment. This leads to degradation of the redshift test accuracy approximately to the level 0.01. We discuss the way to overcome this difficulty and present preliminary results based on data obtained during special observing sessions scheduled for testing the new techniques.}


\section{Introduction}

Phenomenon of  the "photon gravitational redshift" was predicted by Einstein in the paper \cite{E1} and then  discussed in \cite{E2}, i.e. much before the creation of General Relativity (1916). Further it was involved in the number of three famous effects of GR having been got the experimental confirmation \cite{E3}. However as it is accepted now the "photon gravitational redshift" lies in the foundation of GR composing its crucial experimental basis. In particular the redshift effect can be considered as Equivalence Principle Test for photon: i.e. it provides the information concerning the acceleration of photons in gravitational fields \cite{YMT}. Thus a precise measurement of this effect with growing accuracy could define limits of GR validity and stimulate a new physics search.

The Einstein's formulation of the phenomenon consists in the statement that "any clock marches slowly in the gravitational field". So the frequency of atomic clock depends on the value of gravitational potential in the place of its location. The terminology "redshift" has historical origin associated with the first observation of the effect  \cite{Ad} through the measurement of the hydrogen spectral lines in the light coming from the white dwarf Sirius B. 
On the contrary in the Earth gravity field a clock lifted at some altitude has to show a blue shift. Other interpretation of the "electromagnetic gravitational redshift" effect as a loss of the photon energy while traveling through the gravitational field is not completely correct and could lead to contradictions \cite{OST}.

At present time the most precise test of "red shift" effect was performed in the mission of Gravity Probe A (GP-A) of 1976, in which the frequencies of two hydrogen masers clock were compared-one on the Earth and the other on board of a rocket with a ballistic trajectory of $10^4$ km apogee. The experiment \cite{VL} confirmed the value of frequency shift predicted by GR with accuracy $1.4  \times 10^{-4}$. There are several planned experiments aimed at improving the currently achieved accuracy by 2-4 orders of magnitude. The European Space Agency's ACES mission \cite{ACES} intends to install two atomic clocks, an H-maser and the cesium fountain clock (complex PHARAO), onto the International Space Station. The active phase of the mission currently is being scheduled for 2016. Because of the ISS's low orbit, the gravitational potential difference between the ground and the on-board clocks will be only 0.1 of that achievable with a spacecraft at a distance of $\sim 10^5$ km from the Earth. Nevertheless, predicted accuracy, which is expected to be reached $10^{-16}$ in microgravity, provides for measurement accuracy at the level of $10^{-6}$.

Another European initiative is STE-QUEST, a candidate mission for the ESA Cosmic Vision M4 program, with a goal to test RS with $10^{-7}-10^{-8}$ accuracy in the gravitational field of the Earth. Additionally, a special choice of the orbit, which will allow the spacecraft to simultaneously communicate with tracking stations at different continents, will provide opportunities for testing the "redshift" in the field of the Sun. The accuracy of this type of experiment \cite{QEST}, not requiring a frequency standard on board the spacecraft, is speculated to reach $ \sim 10^{-6}$.

Meanwhile the experiment with a potential of testing the "redshift" effect in the field of the Earth with $10^{-6}$ accuracy is currently being carried out as a part of the mission of the space radio telescope (SRT) "Radioastron" (RA) \cite{RA}. The possibility for such measurement came with the decision to add a space hydrogen maser (SHM) frequency standard to the scientific payload of the mission's spacecraft.

However the modes of the high-data-rate radio complex (RDC) at this satellite do not allow independent synchronization of the frequencies of the links used for transmission of tone signals, i.e., 7.2 GHz (up) and 8.4 GHz (down), and the 15 GHz carrier of the data downlink (used for observational and telemetry data transmission). It is possible, however, to independently synchronize the carrier (15 GHz) and the modulation (72 or 18 MHz) frequency of the data downlink. This mixed, or "Semi-coherent," mode of synchronization hasn't been used in astronomical observations so far. As our analysis shows, for this mode it is possible to devise a compensation scheme, which is similar to the one used by the GP-A experiment, and which results in the contributions of the nonrelativistic Doppler effect and the troposphere eliminated in its output signal. The accuracy of the experiment based on this compensation scheme can reach the limit $1.8 \times 10^{-6}$ part of the total gravitational frequency shift, set by the frequency instability and accuracy of the ground and space H-masers (GHM, SHM) \cite{BLR}.

\section{Parameters and operational regimes of RA}

The RA satellite has a very elliptical orbit which changes under the Moon's gravity influence.
Namely: the perigee varies in the region $10^{3} - 80 \times 10^{3}$ km; apogee in $(280 - 350) \times 10^{3}$ km; the orbital period varies in the range 8 -- 10 days; the amplitude modulation of the gravitational frequency shift effect occurs into the interval $(0,4 - 5.8) \times 10^{-10}$. The satellite and land tracking station (Puschino ASC) have in operation equivalent hydrogen frequency standards (production of the national corporation "Vremya-Che") with the following characteristics:
Allan variance under average time $10 - 100$ sec is $3 \times 10^{-14}$,  under $10^{3} - 10^{4}$ sec -- $3 \times 10^{-15}$; the frequency drift was estimated as $10^{-15}$ per day (and $10^{-13}$ per year).
The SHM output signal is transmitted to a TS by the both radio data and radio science complexes (RDC, RSC), which includes two transmitters at 8.4 and 15 GHz, and a 7.2 GHz receiver. The frequencies of the signals used in both complexes, can be synthesized either from the reference signal of the SHM or from the 7.2 GHz output of the on-board receiver, which receives the signal transmitted by the TS and locked to the ground H-maser (GHM). The mode of the on-board hardware synchronization significantly affects not only the achievable accuracy but the very possibility of the gravitational redshift experiment.

In order to compare the output frequencies of a ground $f_{e}$ and a space-borne $f_{s}$ atomic standards, one needs to transmit any (or both) of these signals by means of radio or optical links. The comparison thus becomes complicated by the necessity of extracting
a small gravitational frequency shift from the mix of accompanying effects, such as resulting from the relative motion of the SC and the TS , also the signal propagation through media with non-uniform and time varying refractive indexes. The total frequency shift of a signal, propagating from the SC to the TS, is given by the following equation:
\begin{equation}
f_{\downarrow s} =  f_{s} +\Delta f_{\text{grav}} +\Delta f_{\text{kin}}+ \Delta f_{\text{instr}} + \Delta f_{\text{media}},
\end{equation}
where $f_{ \downarrow s}$ is the frequency of the signal, as received by and measured at the TS, $\Delta f_{\text{grav}}$ is the gravitational frequency shift, $\Delta f_{\text{kin}}$ is the frequency shift due to the SC and TS relative motion, $\Delta f_{\text{media}}$ is the propagation media contribution (ionospheric, tropospheric, interstellar medium), $\Delta f_{\text{instr}}$ encompasses various instrumental effects, which we will not consider here.
The Eq. (1) can be used to determine the gravitational frequency shift $\Delta f_{\text{grav}}$. Indeed, $f_{ \downarrow s}$ is measured at the TS, $\Delta f_{\text{kin}}$ can be evaluated from the orbital data (assuming special relativity is valid), $\Delta f_{\text{media}}$ can be found from multi-frequency measurements (for the ionospheric and interstellar media contributions) and meteorological observations (for the tropospheric one), estimation of $\Delta f_{\text{instr}}$ involves calibration of the hardware and a study of the transmitting, receiving and measuring equipment noise. The value of $f_{s}$ is unobservable and, therefore, presents a certain difficulty. It is convenient to express it in terms of the
frequency of the ground-based standard $f_{e}$ and an offset $\Delta f_{0}$:
\begin{equation}
f_{s} = f_{e} +\Delta f_{0}.
\label{ion}
\end{equation}
The problem of $f_{s}$ (or $\Delta f_{0}$) being not measurable directly has different solutions depending on the type of the frequency standards used and the possibility of varying the gravitational potential difference $\Delta U$ between them.

The two principal but typical operation modes were foreseen and realized in the "Radioastron" satellite so called "H-maser" and "Coherent".

a)	H-maser mode.

This is the main synchronization mode used in radio astronomical observations.
In this mode the SHM signal is used to synchronize both the RDC and RSC frequencies. The frequency $f_{s}$ of the signal, transmitted by the DRC, and frequency 
$f_{ \downarrow s}$ received at a TS are related by the equation (1) where the $\Delta f_{\text{media}}= 
\Delta f_{\text{trop}} + \Delta f_{\text{ion}}$ presented by sum of troposphere and ionosphere shifts. The kinetic shift
having the main contribution from Doppler shift depends on radial velocity of SC in respect of TC $\dot{D}/c = \bf{(v_{s}- v_{e})n}/c$ , where $\bf{n}$ is the unit vector of the view line of the sight.

The kinematic, gravitational and tropospheric contributions are proportional to the transmitted signal frequency, while the ionospheric contribution is inversely proportional to it. The availability of the 2-frequency link ($15$ and $8.4$ GHz) provides for accurate estimation of the ionospheric term, but the contributions of the other effects cannot be separated from each other and need to be calculated from ballistic data. This causes a large error of the
Doppler effect determination and thus degrades the experiment accuracy of the red shift measurement to the value of the order of $1\%$.

b)	Coherent mode.

The "Coherent" mode of the on-board hardware synchronization, also known as the phase-locked loop mode. Here a sinusoidal signal of $ 7.2$ GHz  frequency, synchronized to the GHM, is sent to the SC, where it is received by the RDC and used to lock the frequencies of the RSC. All the signals transmitted to the TS by the RDC are also phase-locked to the received $7.2$ GHz signal: the $8.4$ GHz tone, the $15$ GHz carrier of the data transfer link, and, lastly, its 72 (or 18) MHz modulation frequency.
The "Coherent" mode alone is of no interest to the redshift experiment, because, it is obvious that the received signal has no information about the gravitational redshift effect. However, in the case of simultaneous operation of the one and two-way communication links, their signals can be combined by means of a special radio engineering compensation scheme \cite{VL}, which outputs a signal containing information about the gravitational redshift but, at the same time, free from the 1st-order Doppler and tropospheric contributions. It is also possible to compensate for the ionosphere but only in case of a special selection of the ratios of the up and down link frequencies. Such compensation scheme, first used in the GP-A mission, cannot be applied directly in the case of "RadioAstron," because, the mode of independent synchronization of the carrier frequencies of the RDC links is not supported.

Coming back to the "H-maser mode" let's evaluate a potential sensitivity of the redshift measurement with Radioastron.

During the favorable periods for the gravitational experiment are such that the perigee height is at minimum, (the winter-spring of 2014 and 2016), when it is equal to $1.5 \times 10^{3} $ km, so that $ (U_{ap}-U_{per})/ c^{2} =  5.5 \times 10^{-10}$.The frequency drift of used standards is enough small $1 \times 10^{-15}$ per day -- less then its frequency instability under the $10^{3}$ sec of average time $3 \times 10^{-15}$. Approximately one day is needed for the RA to travel from perigee to a distance where the gravitation potential is almost equal to its value at apogee. Therefore, the accuracy of a single modulation-type experiment is limited not by the frequency drift of either the SHM or the GHM but by the H-maser frequency instability. Thus supposing the compensation of coherent hindrances (Doppler, troposphere, ionosphere), we obtain the following limit for the total relative accuracy of a single experiment to determine the value of the gravitational redshift modulation in the favorable period of low perigee: $5.5 \times 10^{-6}$.

The important advantage of the "RadioAstron" mission, as compared to GP-A, lies in the possibility of conducting the experiment multiple times. Statistical accumulation of measurement results provides reduction of the random error contributions by a factor of $\sqrt{N}$, where $N$ is the number of measurements performed. For $N = 10$, in particular, the contribution of the frequency instability becomes equal to the one of the frequency drift. Since the drift causes a systematic error, accumulating data any further will not improve the experiment accuracy. Then,
we arrive at the following limit for the accuracy of the modulation-type gravitational redshift experiment with "RadioAstron": $1.8 \times 10^{-6}$.

\section{Mixed regime "semi-coherent" and preliminary result}

The new approach to the compensation scheme for Radioastron specific was proposed in the paper \cite{BLR}. 
There is the possibility to synchronize the 8.4 and 15 GHz frequencies of the RDC transmitters to the GHM-locked $7.2$ GHz tone (the "Coherent" mode) of RDC, while the RSC is synchronized to any of the on-board frequency standards. This "Semi-coherent" mode turns out to be the most suitable for the gravitational experiment. Indeed, just like in the "Coherent" mode, the net gravitational redshift is canceled in the received $8.4$ GHz tone and in the carrier of the $15$ GHz data link. However, in contrast to the "Coherent" mode, the modulation frequency of the data link is locked not to the $7.2$ GHz uplink but to the SHM signal, hence all components of the data link signal spectrum, except for the carrier, are influenced by the gravitational effect.
As it is shown in the paper \cite{BLR}, the compensation scheme can be arranged in such a way that all disturbing effects are transferred to the modulation frequency scale, too. Moreover, just like in GP-A, it turns out to be possible to eliminate the contributions of the 1st-order Doppler and tropospheric effects as well. The ionospheric effect is cancelled through the two frequency measurement.

Below we present two pictures as an illustration of the Radioastron operation in the 
"semi Coherent" mode.

\begin{figure}
\includegraphics[width=0.5\linewidth]{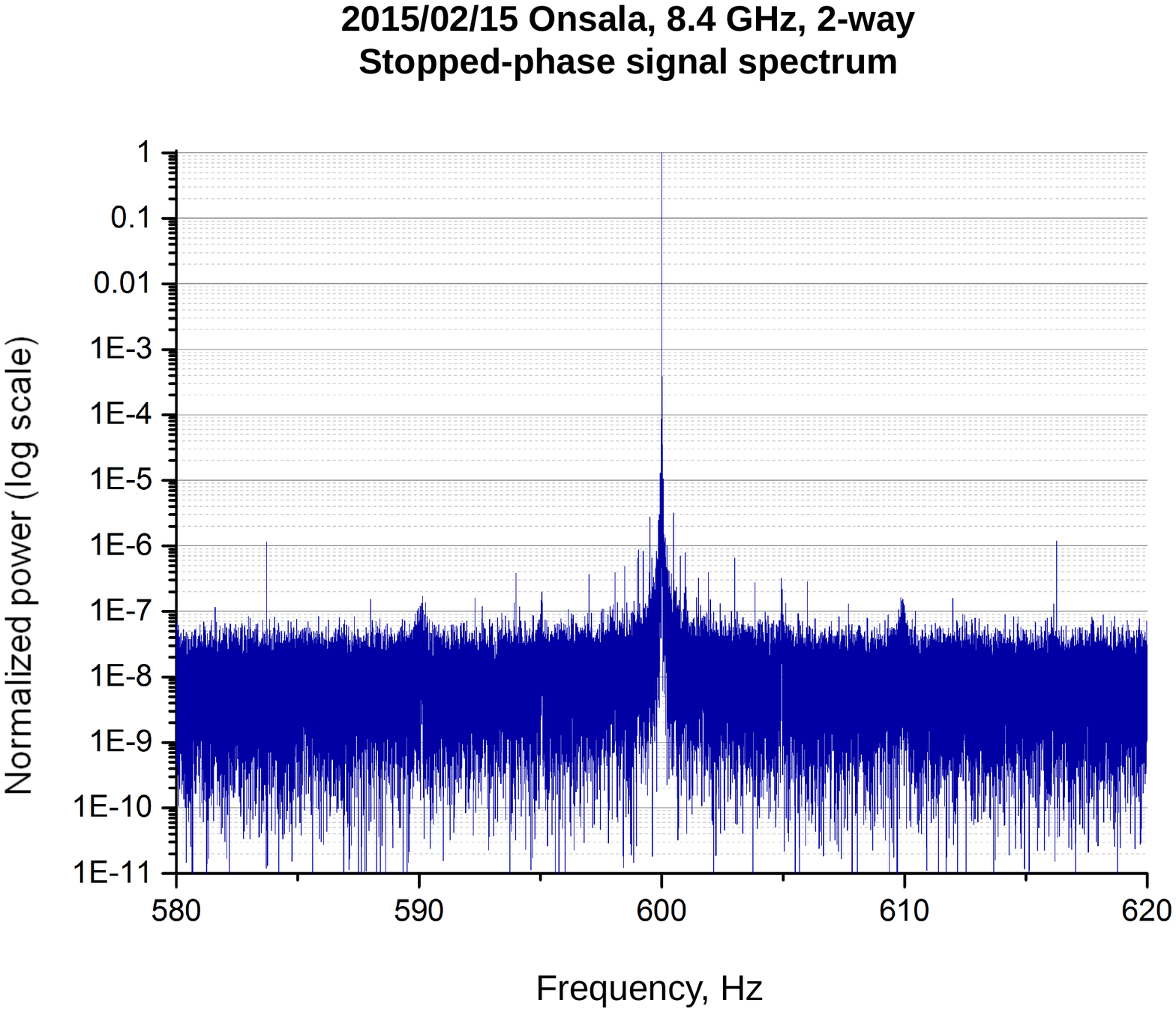}
\hfill
\includegraphics[width=0.5\linewidth]{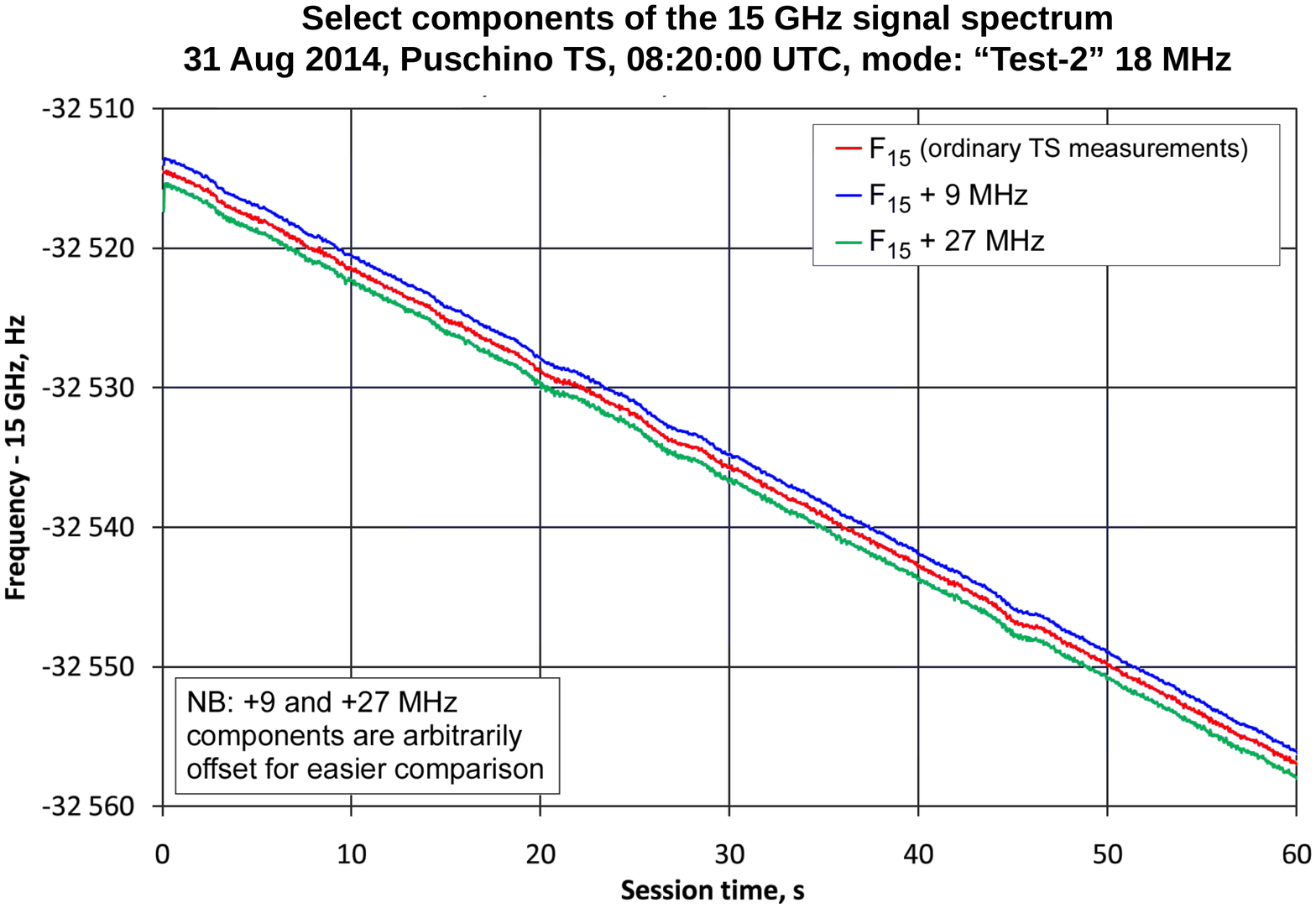}
\caption[]{a) Result of the carrier frequency ($8.4$ GHz) measurement, b) The time evolution of the carrier ($15$ GHz) and two sidebands}
\label{fig:fig1}
\end{figure}

In Fig. \ref{fig:fig1} (a) result of the carrier frequency ($8.4$ GHz) measurement derived from the data received at the Onsala Radio Telescope (Sweden). The width of the line is estimated as $0.0003$ Hz after the complex filtration algorithm used in the radio-astronomical
experiments for moving space apparatus \cite{Mol}. This result provides the accuracy of a single redshift measurement at the level $6 \times 10^{-5}$ (the absolute magnitude of the red shift is $5$ Hz).

At the Fig. \ref{fig:fig1} (b) the time evolution of the carrier ($15$ GHz) and two sidebands is presented as it was recorded at the Puschino tracking station. It demonstrates the similar (coherent) time perturbations for the central line and sidebands, despite of the fact they were synchronized from different 
(land and board) standards.

In the final remark we would like to name the principle Institutions in Russia performing hard work of accompany the Radioastron mission; in that number are: the Astro Space Center of the Lebedev Physical Institute RAS, Sternberg Astronomical Institute MSU, Keldysh Institute for Applied Mathematics RAS and Lavochkin Scientific and Production Association. There are also several world Institutions interested in gravitational program of the mission, they are: The
York University (Canada) ( N. Bartel, W. Cannon), Joint Institute for VLBI in Europe (Netherlands) (L. Gurvits, S. Pogrebenko, G. Cimo ), University of California in Santa-Barbara (USA) (C. Gwinn, M. Johnson), Hartebeesthoek Radio Observatory (South Africa) (M. Bietenholz). Periodically the observational service is carried out by the radio telescope observatories in Green Bank (USA), Effelsberg (Germany), Onsala (Sweden) and others.

\section*{Acknowledgments}

Authors would like gratitude the main ideologist and leader of the Radioastron mission
Nikolai Kardashev (director ASC) for permanent attention and assistance to the gravitational
group of the scientific program RA.

\end{document}